\documentclass[epj]{webofc}
\usepackage[varg]{txfonts}   
\graphicspath{{figs/}}
\begin{document}
\title{Pion transition form factor in the Domain model of QCD vacuum}

\author{\firstname{Sergei} \lastname{Nedelko}\inst{1}\fnsep
\thanks{\email{nedelko@theor.jinr.ru}} \and
        \firstname{Vladimir} \lastname{Voronin}\inst{1}
}

\institute{Bogoliubov Laboratory of Theoretical Physics, JINR,
141980 Dubna, Russia
}

\abstract{
Domain model of QCD allows for description of wide range of meson observables, dynamical chiral symmetry breaking, and resolution of $U_A(1)$ and strong $CP$ problems. The purpose of the present study is transition form-factor of neutral pseudoscalar mesons, and how they are influenced by the typical vacuum configurations of the model -- almost everywhere homogeneous Abelian (anti-)self-dual fields. Asymptotic behavior of the calculated pion form-factor supports Belle trend, but the asymptotic value of $Q^2F_{\pi\gamma^*\gamma}$ differs from the prediction of factorization theorems.
}
\maketitle
\section{Introduction}
Experimental data for pion transition form-factor $F_{\pi\gamma^*\gamma}$ obtained by the BaBar collaboration~\cite{Aubert:2009mc} indicate growth of $Q^2 F_{\pi\gamma^*\gamma}$ at large $Q^2$ that is inconsistent with rigid prediction of QCD factorization theorems~\cite{Lepage:1980fj}
\begin{equation}
\label{form-factor_factorization_asym}
F_{\pi\gamma^*\gamma}\sim \frac{\sqrt{2} f_\pi}{Q^2},
\end{equation}
where $f_\pi=131\ \mathrm{MeV}$. Published later results of experiments on pion transition form-factor $F_{\pi\gamma^*\gamma}$ carried out by the Belle~\cite{Uehara:2012ag} collaboration demonstrate qualitatively different behavior at large momenta, though they still allow violation of bound~\eqref{form-factor_factorization_asym}.

Particular source of contributions that could explain BaBar trend is the QCD vacuum described by ensemble of almost everywhere homogeneous Abelian (anti-)self-dual fields with nonzero scalar condensate $\langle F^2 \rangle$ --- the key feature of the domain model~\cite{EN1,EN,NK1,NK4,Nedelko:2016gdk} --- that can possibly mix short and long-range dynamics. Short-range fluctuations in the model are separated from long-range modes with scalar condensate and integrated out, while the latter are treated nonperturbatively during bosonization procedure. Chiral symmetry is spontaneously broken by vacuum field, there exists nonzero scalar quark condensate $\langle \bar{q} q\rangle$ (see~\cite{Nedelko:2016gdk} for details). The $U_A(1)$ problem is resolved without introducing the strong charge-parity ($CP$) violation. This approach demonstrated its power in description of wide range of meson phenomenology (masses of light, heavy-light and double-heavy mesons, leptonic decay constants, transition constants, all of the above including excited mesons). Overall precision of the approach in the lowest approximation is about $10-15\%$ with a few exceptions. The model also allows consistent treatment of processes involving higher number of hadrons. Results of calculations of decay constants $g_{VPP}$ are given below.

Propagators and meson-quark vertices in the vacuum field are not translation-invariant, therefore momentum is not conserved in every vertex but only in an entire diagram. The purpose of the present paper is to investigate how this feature of the domain model  affects transition form-factors of pseudoscalar mesons. All calculations are performed with the same set of parameters that was used for calculation of meson spectra~\cite{Nedelko:2016gdk}, see Table~\ref{values_of_parameters}.
\begin{table}
\begin{tabular}{@{}ccccccc@{}}
\hline\hline
$m_{u/d}$(MeV)&$m_s$(MeV)&$m_c$(MeV)&$m_b$(MeV)&$\Lambda$(MeV)&$\alpha_s$&$R$(fm)\\
\hline
$145$&$376$&$1566$&$4879$&$416$&$3.45$&$1.12$\\
\hline\hline
\end{tabular}
\caption{Values of parameters used for calculations.
\label{values_of_parameters}}
\end{table}

\section{Decay constants and transition form-factor of pion\label{section_formfactors}}
\begin{figure}
{\centering
\parbox{0.26\textwidth}{\includegraphics[scale=0.3]{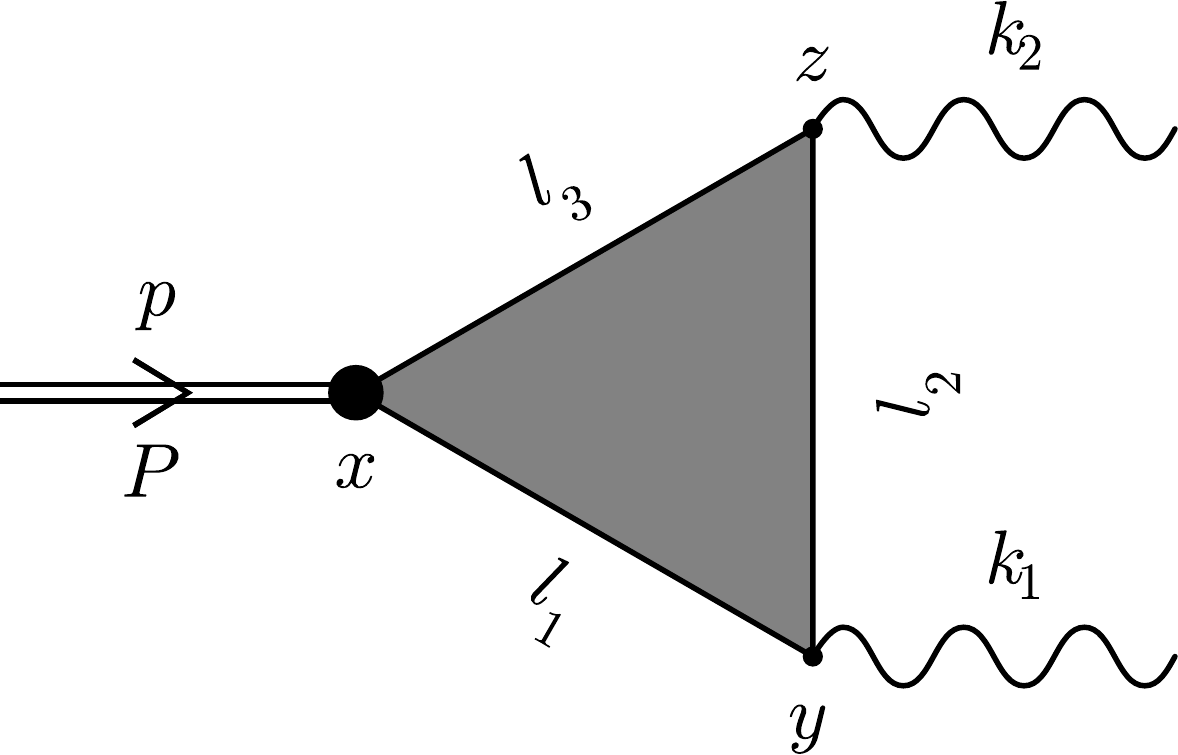}}
\parbox{0.22\textwidth}{\includegraphics[scale=0.3]{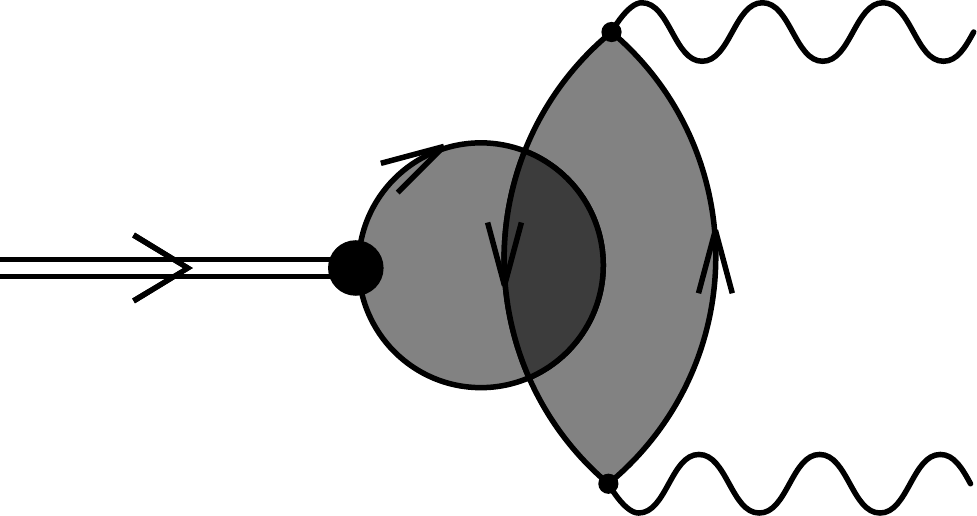}}
\parbox[c][\totalheight][t]{0.42\textwidth}{
\raisebox{-\height}{\includegraphics[scale=0.3]{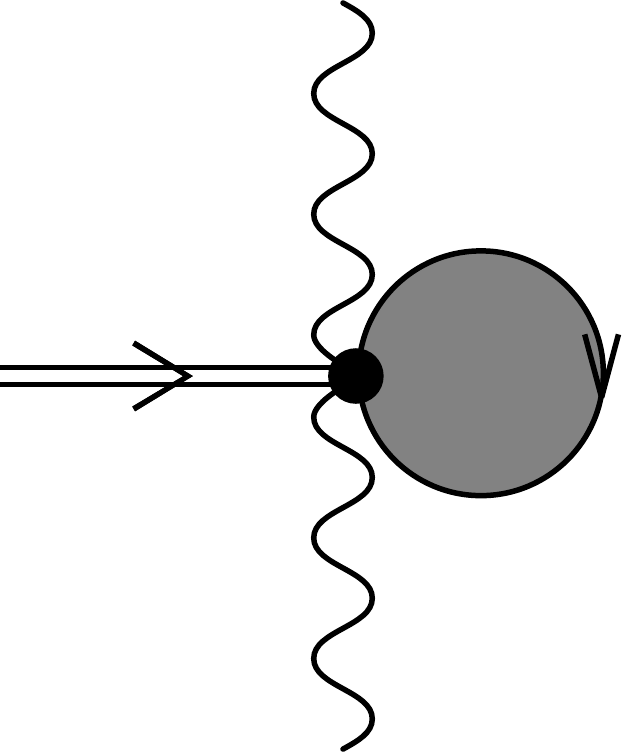}}
\hspace*{0.3em}
\raisebox{-\height}{\includegraphics[scale=0.3]{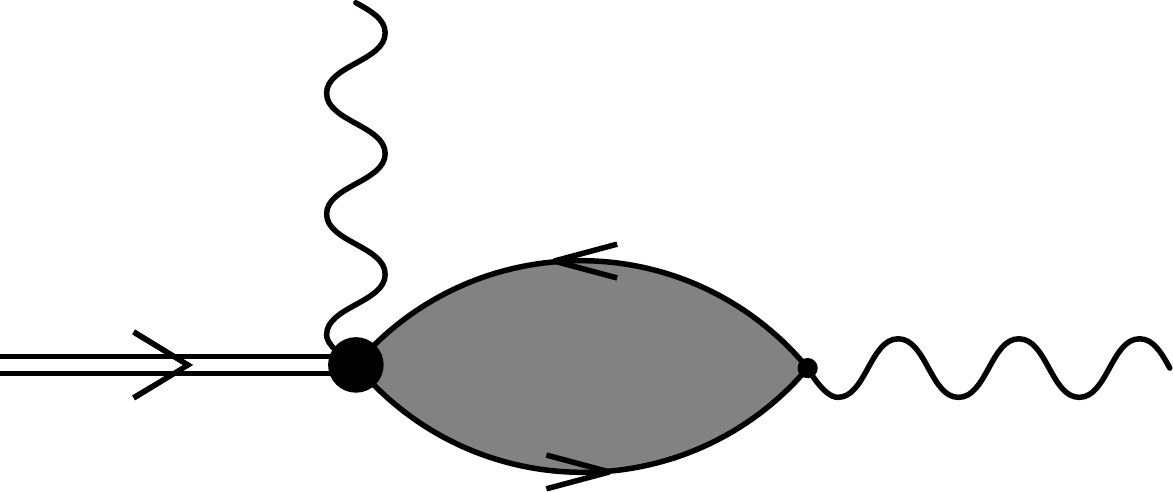}}
}}\\[1em]
\parbox{0.26\textwidth}{\centering A}
\parbox{0.22\textwidth}{\centering B}
\parbox{0.13\textwidth}{\centering C}
\hspace*{0.3em}
\parbox{0.24\textwidth}{\centering D}
\caption{Possible contributions to transition form factor. \label{F_P_gamma_picture}}
\end{figure}

\begin{figure}
\sidecaption
\includegraphics[width=0.49\textwidth]{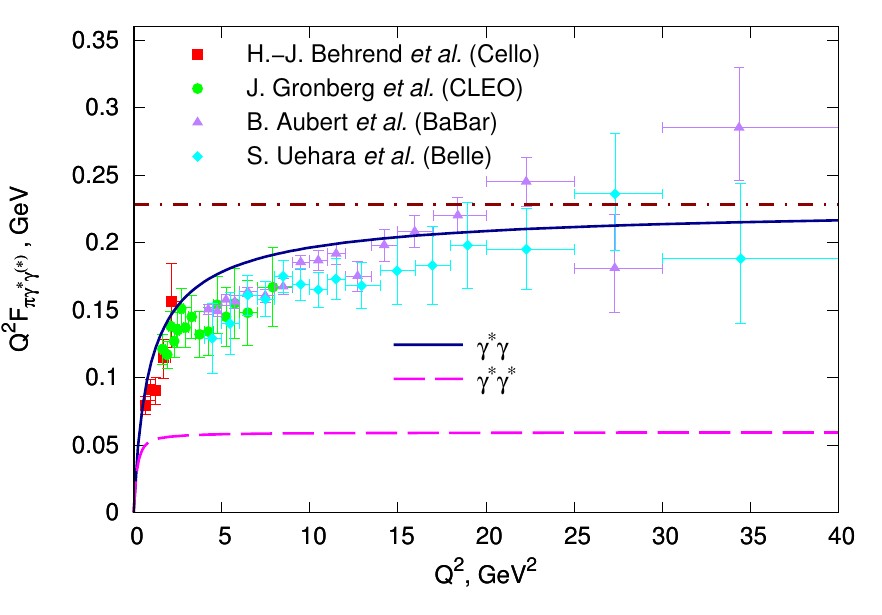}
\caption{Pion transition form-factor in asymmetric (solid line) and symmetric (dashed line) kinematics. $g_{\pi\gamma\gamma}=F_{\pi\gamma^*\gamma(0)}=0.272\ \mathrm{GeV}^{-1}$. The data are taken from \cite{Behrend:1990sr,Gronberg:1997fj,Aubert:2009mc,Uehara:2012ag}.
\label{pi_transition_figure}}
\end{figure}

Interaction of a meson with two photons is described by the diagrams shown in Fig. \ref{F_P_gamma_picture}. Currently, the inhomogeneity of the vacuum gluon ensemble is not taken into account completely. As a consequence, only the diagram A gives nonzero contribution.
\setlength{\fboxsep}{0pt}
Result of computation of pion transition form-factor in comparison with experimental data is shown in Fig. \ref{pi_transition_figure}.

It turns out that nonconservation of momentum in every single vertex does not cause growth of $Q^2 F_{\pi\gamma^*\gamma}$ at large $Q^2$, and calculations show that at asymptotically large $Q^2$
\begin{equation*}
F_{\pi\gamma^*\gamma}\sim \varkappa\frac{\sqrt{2}f_\pi}{Q^2},\quad \varkappa\approx 1.24.
\end{equation*}
Qualitatively, asymtotic behavior of $Q^2 F_{\pi\gamma^*\gamma}$ at large $Q^2$ coincides with the factorization prediction, but the value of constant $\varkappa$ substantially differs from unity. This difference is a result of translational noninvariance of vertices. At the same time, this noninvariance does not affect asymptotic behavior of form-factor in symmetric kinematics (two photons with equal virtuality $Q^2$). Thus, asymptotics calculated within the model
\begin{equation*}
F_{\pi\gamma^*\gamma^*}\sim \varkappa^*\frac{\sqrt{2}f_\pi}{3Q^2},\quad \varkappa^*=1
\end{equation*}
matches factorization prediction
\begin{equation}
\label{form-factor_factorization_sym}
F_{\pi\gamma^*\gamma^*}\sim \frac{\sqrt{2} f_\pi}{3Q^2}
\end{equation}

The fact that prediction~\eqref{form-factor_factorization_sym} is reproduced by the model, while prediction~\eqref{form-factor_factorization_asym} is not, comes as no surprise because straightforward QCD factorization works well in the former case only, i.e.~when both photons are highly virtual~\cite{Radyushkin:1996tb,Mikhailov:2009kf}.

\section{Strong decays of vector mesons\label{section_vpp_decays}}
\begin{figure}
\sidecaption
\includegraphics[scale=0.35]{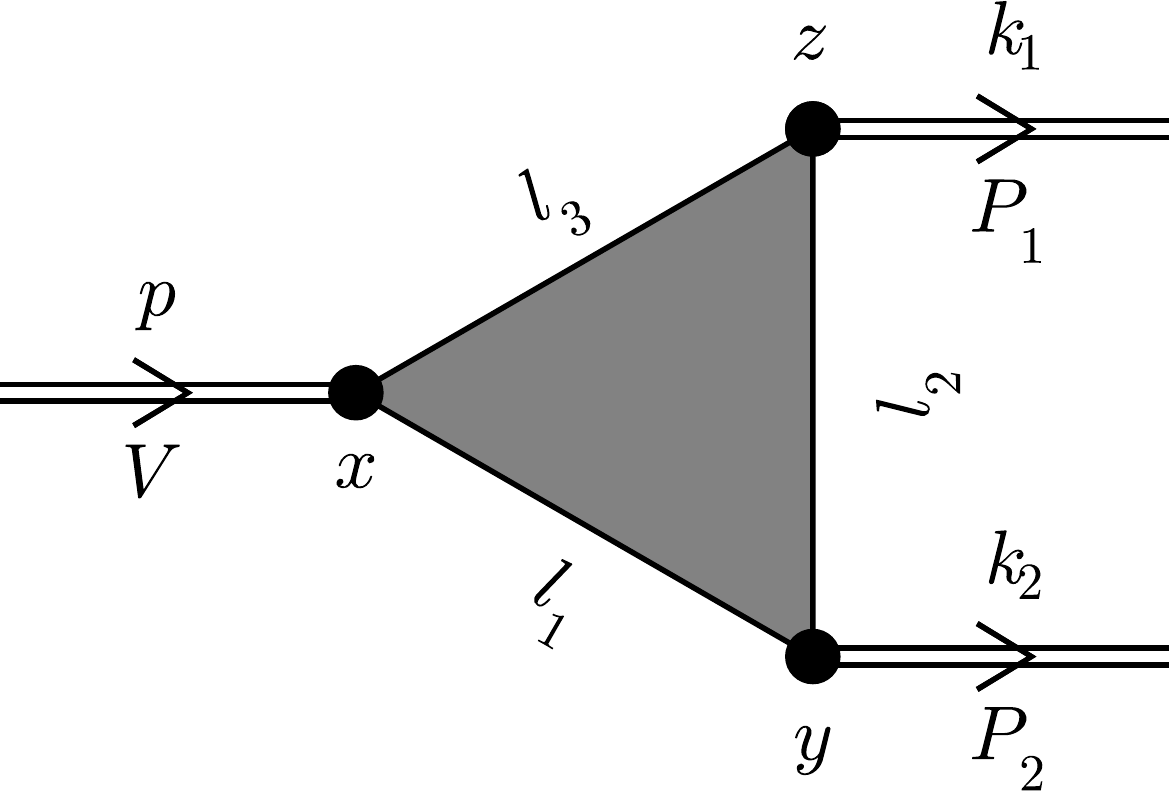}
\caption{Diagram of decay of vector meson into a couple of pseudoscalar mesons.} \label{VPP_decay_diagram}
\end{figure}

The diagram describing process of decay of vector meson into a couple of pseudoscalar mesons is shown in Fig.~\ref{VPP_decay_diagram}. This process is treated consistently with previously obtained within the model meson observables. Calculated values of $g_{VPP}$ are given in Table~\ref{table_vpp_decays}.
\begin{table}
\begin{tabular}{|c|c|c|c|}
\hline
Decay & $g_{VPP}$ \cite{PDG} & $g_{VPP}$&$g^*_{VPP}$\\
\hline
$\rho^0\rightarrow \pi^+ \pi^-$ & $5.95$ & $7.61$&$1.14$\\
\hline
$\omega\rightarrow \pi^+ \pi^-$ & $0.17$ & $0$&$0$\\
\hline
$K^{*\pm} \rightarrow K^\pm \pi^0$ & $3.23$ &$3.56$&$0.65$\\
\hline
$K^{*\pm} \rightarrow K^0 \pi^\pm$ & $4.57$ &$5.03$&$0.91$\\
\hline
$\varphi\rightarrow K^+ K^-$ & $4.47$ &$5.69$&$1.11$\\
\hline
$D^{*\pm}\rightarrow D^0 \pi^\pm$ & $8.41$ &$7.94$&$16.31$\\
\hline
$D^{*\pm}\rightarrow D^\pm \pi^0$ & $5.66$ &$5.62$&$11.53$\\
\hline
\end{tabular}
\caption{Results of calculation of $g_{VPP}$ for various decays. $g^*_{VPP}$ are results that one would obtain if local gauge invariance was neglected. Parameters of the model are given in Table~\ref{values_of_parameters}. \label{table_vpp_decays}}
\end{table}

Invariance of meson-meson amplitude under local background field gauge transformation turns out to be of crucial importance for correct description of the decays. If one reduces local gauge invariance to just the global one, decay constants  change dramatically, especially $g_{\rho\pi\pi}$ (compare third and fourth column of Table~\ref{table_vpp_decays}). This is to be compared with usually underestimated $g_{\rho\pi\pi}$ decay constant~\cite{Bernard:1993wf,Deng:2013uca}. Value of $g_{\omega\pi\pi}$ is exactly zero because of ideal mixing of $\omega$ and $\phi$ mesons and employed approximation of $SU(2)$ isospin symmetry ($m_u=m_d$).

\section{Outlook\label{section_outlook}}
It is important to describe transition form-factors of $\eta,\eta',\eta_c$ simultaneously with pion. It was shown that in the instanton liquid model diagrams of type B in Fig.~\ref{F_P_gamma_picture} contribute to transition form-factors~\cite{Kochelev:2009nz}. While this contribution is suppressed by difference of masses of up and down quarks and negligible in the present consideration, it should be taken into account in order to describe transition form-factors of other mesons. However, incorporating this contribution into the domain model requires explicit accounting for inhomogeneity of the vacuum ensemble.

\end{document}